\documentclass[a4paper]{article}

\usepackage{graphics} % for pdf, bitmapped graphics files
\usepackage{epsfig} % for postscript graphics files
\usepackage{amsmath} % assumes amsmath package installed
\usepackage{amssymb}  % assumes amsmath package installed
\usepackage{algorithm} % to use algortihm layout
\usepackage{algpseudocode} % part of algorithmicx
\usepackage{subcaption}
\usepackage{multirow}
\usepackage{caption}
\usepackage{graphicx}

\title{\LARGE \bf
An Exponentially Stable Extended Kalman Filter with Estimate-dependent Process noise Covariance for Chemical Reaction Networks
}

\author{Suryasnata Dash$^{1}$, Abhishek Dey$^{1}$% <-this % stops a space
\thanks{$^{1}$Suryasnata Dash and Abhishek Dey are with the Department of Electrical Engineering,
        National Institute of Technology, Rourkela, India
        {\tt\small 523ee1001@nitrkl.ac.in, deyab@nitrkl.ac.in}}%
}

\begin{document}

\maketitle
\thispagestyle{empty}
\pagestyle{empty}

%%%%%%%%%%%%%%%%%%%%%%%%%%%%%%%%%%%%%%%%%%%%%%%%%%%%%%%%%%%%%%%%%%%%%%%%%%%%%%%%
\begin{abstract}
Biomolecular systems are often modeled with partially known nonlinear stochastic dynamics, making state and parameter estimation a central challenge. While Kalman filtering techniques are widely used in this setting, their performance critically depends on the choice of the process noise covariance, which is typically assumed constant and heuristically tuned. Such assumptions are not justified for biomolecular systems, where intrinsic noise arises from underlying reaction kinetics. In previous works, a process noise covariance update based on the Chemical Langevin Equation (CLE) was introduced for Extended Kalman Filter (EKF)-based estimation in Chemical Reaction Networks (CRN). In this work, we analyze the stochastic stability of this filtering framework. In particular, we obtain a conservative upper bound on sampling interval for discrete-time biomolecular systems that ensures mean-square exponential boundedness under stated assumptions. The proposed framework is validated through simulations on a nonlinear gene expression model. The analysis provides theoretical justification for CLE-based process noise covariance modeling in EKF design for biomolecular circuits, reducing reliance on heuristic covariance tuning.
\end{abstract}

%%%%%%%%%%%%%%%%%%%%%%%%%%%%%%%%%%%%%%%%%%%%%%%%%%%%%%%%%%%%%%%%%%%%%%%%%%%%%%%%
\section{INTRODUCTION}
Recent advances in systems and synthetic biology have enabled the application of engineered biomolecular circuits in various fields such as agriculture, bioremediation and biofuel generation. Mathematical modeling is the fundamental tool to achieve predictable and scalable performance in these engineered systems. Due to the partial knowledge of the states and parameters, estimation is often one of the main objectives in the analysis of biomolecular systems. The inherent nonlinearity and stochasticity of these systems make the estimation problem challenging.

Biomolecular systems are typically modeled as Chemical Reaction Networks (CRN) either in the deterministic or stochastic sense~\cite{del2015biomolecular}. In the stochastic setting the time evolution of CRNs are described by Chemical Master Equations (CME). However, CMEs are often analytically and computationally intractable for most CRNs due to their high dimension. Alternatively Chemical Langevin Equations (CLE) represents an approximate reaction dynamics as stochastic differential equations, or difference equations in case of discrete time representation~\cite{gillespie2000chemical}. The state-dependent diffusion term in the CLEs represent the intrinsic noise in biomolecular systems~\cite{mcadams1999sa,thattai2001intrinsic} that occur due to uncertainty in chemical reactions. Since CLE diffusion terms explicitly characterize intrinsic stochasticity, they naturally provide a first principle-based candidate for process noise covariance modeling in recursive Bayesian estimation techniques.

Kalman filters and its variants are one of the widely used filtering techniques for state and parameter estimation in any physical system. These are recursive state estimators where optimal gain is assigned minimizing filter error and has been used for estimation in systems biology. Extended Kalman filter (EKF) along with a posteriori identifiability check has been used for parameter estimation and model selection in biomolecular systems~\cite{lillacci2010parameter}. State and parameter estimation for heat shock response models in microbes using EKF, unscented Kalman filter (UKF), and particle filters (PF) are compared by Liu et al.~\cite{liu2012state}. Kurdyaeva et al. have used EKF to estimate Monte Carlo-based higher order moments to study time evolution of moments in nonlinear biochemical networks~\cite{kurdyaeva2021uncertainty}. In~\cite{linden2022bayesian} the authors have used a constrained interval UKF (CIUKF) with Markov Chain Monte Carlo (MCMC) sampling to estimate parameters in systems biology models. The authors in~\cite{abolmasoumi2022robust} have proposed a robust generalized maximum likelihood type UKF for state estimation in gene regulatory networks. While these studies provide important groundwork for state and parameter estimation in CRNs, rigorous stability analysis of Bayesian filtering methods for state estimation in such systems remains largely unexplored.

 Filter stability is a necessary condition for reliable estimate of the unknown states and parameters. For stochastic dynamical systems, such as CRNs, filter stability must be analyzed in a probabilistic framework, typically formulated through moment-based criteria such as boundedness of the mean squared estimation error. Stochastic stability ensures convergence and boundedness of filter error with time. Literature shows significant studies of filter stability using Lyapunov function. Stochastic stability defined based on condition of Lyapunov function has been explained in~\cite{kushnerstochastic} and~\cite{karvonen2014stability}. It describes the bounded nature of the process $e_k$ with initial guess $e_0$ and a constant $C_0$, 
 \begin{equation*}
 	E(\|e_k\|^2) < E(\|e_0\|^2) + C_0
 \end{equation*}  

Deyst et al.~\cite{deyst2003conditions, deyst2003correction}, and Jazwinski~\cite{jazwinski2007stochastic} have elaborated on filter achieving asymptotic stability under certain assumptions, but one of its main assumption considered the system matrix to be invertible which is difficult to implement in practical scenarios. Moore et al.~\cite{moore1980coping} discussed on the exponential stability along with relaxation of the assumption of system matrix being invertible. Reif et al.~\cite{reif1999stochastic} studied the exponential boundedness of EKF, but it also considered the system matrix to be invertible. Karvonen et al.~\cite{karvonen2020stability} discussed continuous time EKF stability and has relaxed the former assumption. Wei et al.~\cite{wei2025stochastic} validated the exponential boundedness of discrete system based filter, with Lipschitz continuity condition of the state transformation function instead of invertibility assumption. Existing stochastic stability analyses of EKF-type filters generally assume fixed or uniformly bounded process noise covariance matrices independent of the filter estimates. In contrast, if the covariance update depends explicitly on the estimated state then how to analyze stochastic stability of such a filter is generally unclear. 

In literature the process noise covariance matrix for biomolecular systems is typically considered to be constant and decided by trial and error based on how much the process model can be trusted~\cite{lillacci2010parameter,sun2008extended}. To mitigate this, in our previous work, we have proposed a process noise covariance matrix for CRNs which is updated at each instant based on the state-dependent diffusion terms in the CLE model~\cite{dash2025extended}. Preliminary investigations show that this choice of process noise covariance matrix can yield improved innovation whiteness characteristics and can achieve a balance between mean squared estimation error and estimation convergence time~\cite{dash2025extended,dey2019kalman}. EKF with updating process noise covariance matrix is claimed to be the best linear unbiased estimator as compared to EKF with fixed process noise covariance matrix~\cite{gola2025kalman}, but stochastic stability for the same has not been explained. The dependence of covariance update through the CLE diffusion term introduces additional coupling between the estimation error dynamics and error covariance evolution making stochastic stability analysis of such a filter non-trivial. Motivated by this, we aimed to analyze the stochastic stability of EKF with such a process noise covariance matrix that is dependent on state estimates of previous instant. The main contribution of this letter is the analysis of mean-square exponential boundedness of an EKF with estimate-dependent process noise covariance derived from CLE diffusion terms. The analysis yields a conservative upper bound on the sampling interval that guarantees the boundedness of the estimation error. Section \MakeUppercase{\romannumeral2} describes the system and construction of nonlinear KF. Section \MakeUppercase{\romannumeral3} explains the analysis for filter stability. Section \MakeUppercase{\romannumeral4} validates the proof with results of biomolecular system example and Section \MakeUppercase{\romannumeral5} summarizes the overall work.

\section{SYSTEM DESCRIPTION AND DEFINITIONS}
\subsection{Dynamics of Biomolecular system}
A discrete stochastic biomolecular system can be represented as,
\begin{gather*}
	\vspace{-5pt}
	x_{k+1} = f(x_k) + G_kw_k \\
	y_k = Cx_k + v_k, \tag{1} \label{nl-eqn}
\end{gather*}
where $f:\mathbb{R}^{n\times 1} \rightarrow \mathbb{R}^{n\times 1}$ is a nonlinear function, $x \in \mathbb{R}^{n\times 1}$ is species of the system with $n$ denoting number of species, $G_k \in \mathbb{R}^{n\times m}$ is noise coefficient matrix with $m$ indicating the number of reactions occurring. The measurements, $y_k \in \mathbb{R}^{p\times 1}$ are taken as linear function of the states. $w_k \in \mathbb{R}^{m\times 1}$ and $v_k \in \mathbb{R}^{p\times 1}$ are Gaussian process and measurement noise where $w_k \sim \mathcal{N}(0,Q_0)$ and $v_k \sim \mathcal{N}(0,R)$. The process dynamics in~\eqref{nl-eqn} is analogous to CLE~\cite{gillespie2000chemical}, 
\[ x_{k+1} = x_k + \delta V\mathcal{A}(x_k) + G_kw_k. \tag{2} \label{cle-eqn}\]
Here, $V$ is the stochiometric matrix of $\mathbb{R}^{n\times m}$ and $\mathcal{A}(x_k)$ is the propensity vector of $\mathbb{R}^{m\times 1}$ consisting propensities $a_{1k}, a_{2k}, \dots, a_{mk}$. The diffusion coefficient $G_k = \sqrt{\delta}V\tilde{\mathcal{A}}(x_k)$, where  \[\tilde{\mathcal{A}}(x_k) = 
\begin{bmatrix} 
	\sqrt{a_{1k}} & 0 & \cdots & 0 \\
	0  & \sqrt{a_{2k}} & \cdots & 0 \\
	\vdots & \vdots & \ddots & \vdots \\
	0 & 0 & \cdots & \sqrt{a_{mk}}
\end{bmatrix}.
\]

\subsection{State Estimation with EKF}
For nonlinear systems, EKF is a common estimation algorithm where concept of Jacobian matrix is used to compute the error covariance matrix. The EKF process is split into prediction and correction step. The equations involved are shown in Algorithm~\ref{alg:ekf}~\cite{simon2006optimal}.

\begin{algorithm}[thbp]
	\caption{EKF Process}
	\label{alg:ekf}
	\begin{algorithmic}
		\Require System from~\eqref{nl-eqn} with $w_k \sim \mathcal{N}(0,Q_0)$ where $Q_0 = I_{m\times m}$ and $v_k \sim \mathcal{N}(0,R)$
		\Statex Prior information: $\hat{x}_0^+$,$P_0^+$
		\Statex Measurements: $y = \{y_1, y_2 \dots y_N\}$
		\For{$k=1,2,\dots,N$}
		\State \textbf{Prediction}:
		\State $\hat{x}_{k+1}^- = f(\hat{x}_{k}^+)$
		\State $P_{k+1}^- = F_{k}P_{k}^+F_{k}^T + Q_{k}$ \State where $F_{k} = \frac{\partial f(\hat{x}_{k}^+)}{\partial \hat{x}_{k}^+}$, $Q_{k} = G_{k}Q_0G_{k}^T$
		\State \textbf{Correction}:
		\State $\hat{x}_{k+1}^+ = \hat{x}_{k+1}^- + K_{k+1}(y_{k+1} - C\hat{x}_{k+1}^-)$
		\State $P_{k+1}^+ = (I - K_{k+1}C)P_{k+1}^- $ %(I-K_{k+1}C)^T +$ %\State $K_{k+1}RK_{k+1}^T$ 
		\State where $K_{k+1} = P_{k+1}^-C^T(CP_{k+1}^-C^T + R)^{-1}$
		\EndFor
	\end{algorithmic}
\end{algorithm}

In our proposed filter, $Q_k$ in Algorithm~\ref{alg:ekf} is, 
\begin{equation*}
	Q_k = G_kG_k^T = \delta V\tilde{\mathcal{A}}(\hat{x}_k^+)\tilde{\mathcal{A}}(\hat{x}^+_k)^TV^T. \tag{3} \label{updating_Q}
\end{equation*}
This represents the dependence of process noise covariance matrix on the previous state estimates. We analyze the stability of EKF with such a process noise covariance dependent on state estimates.

\section{STOCHASTIC STABILITY ANALYSIS}
Stochastic stability involves the study of system stability where noise exists and instead of system converging to equilibrium state, it is bounded within certain limits. The basic definition of stochastic stability in exponential sense is mentioned in~\cite{1101300} and further explained in~\cite{karvonen2014stability}: 
\vspace{6pt} \newline
\textbf{Definition 1}: \textit{A stochastic process $e_k$ is exponentially bounded in mean square if \[E(\|e_k\|^2) \le C_1\gamma^kE(\|e_0\|^2) + C_0,\] where $C_1,C_0$ are real numbers with $C_1,C_0 > 0$ and $0<\gamma<1$ $\forall \ k$.} \newline
Similar to approach of validating stability based on Lyapunov functions in deterministic systems~\cite{khalil2002nonlinear}, literature also focused on using Lyapunov functions that satisfy supermartingale logic~\cite{BUCY1965151} to verify the stochastic stability of the system~\cite{1101300}. However it is mentioned as a sufficient condition in the same, and we analyze stochastic stability of the proposed filter based  on Definition 1. 

The filter stability is validated considering certain assumptions: \newline
\textbf{Assumption 1.} Process noise covariance (Q) is positive semi-definite and measurement noise covariance (R) is positive definite with known covariance $\|R\| = r$. \newline
\textbf{Assumption 2.} The system considered is stable and as assumed in~\cite{karvonen2020stability}, the drift term $f(x_k)$ is locally Lipschitz i.e, $ \|f(x_{k}) - f(x_{k'})\| \le L_f\|x_k - x_{k'} \|$ where $L_f < 1$. \newline
\textbf{Assumption 3.} Prior information error $x_0 - \hat{x}_0$ is known and uncorrelated with process and measurement noise $w_k$ and $v_k$ respectively. \newline
\textbf{Assumption 4.} The propensities are locally Lipschitz i.e, $ \| \mathcal{A}(x_k) - \mathcal{A}(x_{k'})\| \le L_a\|x_k - x_{k'} \|$ with $0 < L_a < \infty$. \newline
The main contribution of our work is stated in the following theorem,\\
\textbf{Theorem 1:} \textit{
	If from Definition 1, $\gamma = g(\delta)$ where $g:\mathbb{R} \rightarrow \mathbb{R}$ and there exists a $\delta$ such that \[\delta < \delta_{max},\]
	 where $\delta_{max}$ is a conservative bound, then filter error $e_k$ of an EKF with updating process noise covariance based on CLE in~\eqref{updating_Q} satisfies Definition 1.
}

To prove the above theorem, we follow the sequential steps: 
\begin{itemize}
	\item[1.] We find $\|G_k\|^2$.
	\item[2.] We find $\|P^+_{k+1}\|$, $\|K_{k+1}\|$ and $\|I - K_{k+1}C\|$ based on result obtained in Step 1.
	\item[3.] We find $E(\|e_{k+1}\|^2 | e_k)$ and then find $E(\|e_{k+1}\|^2)$ based on previous steps.
	\item[4.] We check whether $E(\|e_{k+1}\|^2)$ satisfies the stochastic stability definition and based on what condition we can ensure the stochastic stability of the filter.
\end{itemize} \vspace*{5pt}
\textbf{Step 1:} Our first step is to derive the bound of Q based on CLE. We find the Euclidean norm, $\|Q_k\| = \|G_kG_k^T\| \le \|G_k\|^2$. Using Theorem 1.3.22 in~\cite{horn2012matrix}, $\lambda_{max}(G_k^TG_k) = \lambda_{max}(G_kG_k^T)$ and,
\begin{equation*} 
	\|G_k\| = \sqrt{\lambda_{max}(G_kG_k^T)} \implies \|G_k\|^2 = \lambda_{max}(G_kG_k^T). 
\end{equation*}
We use $G_kG_k^T$ defined in~\eqref{updating_Q}:
 \begin{align*}
	\|G_k\|^2 =\delta\lambda_{max}(V\tilde{\mathcal{A}}(\hat{x}^+_k)\tilde{\mathcal{A}}(\hat{x}^+_k)^TV^T), \tag{4} \label{ineq1}
\end{align*} 
Since matrix in~\eqref{updating_Q} is a symmetric: $$\|V\tilde{\mathcal{A}}(\hat{x}^+_k)\tilde{\mathcal{A}}(\hat{x}^+_k)^TV^T\| = \lambda_{max}(V\tilde{\mathcal{A}}(\hat{x}^+_k)\tilde{\mathcal{A}}(\hat{x}^+_k)^TV^T).$$
$\tilde{\mathcal{A}}(\hat{x}^+_k)$ is a diagonal matrix hence $\tilde{\mathcal{A}}(\hat{x}^+_k)\tilde{\mathcal{A}}(\hat{x}^+_k)^T = \tilde{\mathcal{A}}_k^2$. Since $V$ is dependent on system itself and known, $\|V\| \le \mathbf{v}$. Implementing the norm equality, 
\begin{gather*}
\|V\tilde{\mathcal{A}}(\hat{x}^+_k)\tilde{\mathcal{A}}(\hat{x}^+_k)^TV^T\| \le \|V\|\|\tilde{\mathcal{A}}(\hat{x}^+_k)^2\|\|V^T\| \le \mathbf{v}^2\|\tilde{\mathcal{A}}(\hat{x}^+_k)^2\|.
\end{gather*} 
A significant observation is that, $\lambda_{max}(\tilde{\mathcal{A}}(\hat{x}^+_k)^2)$ is $max\{a_{1k},a_{2k} \dots a_{mk}\}$ meaning:\newline
$$\|\tilde{\mathcal{A}}(\hat{x}^+_k)^2\| < \|\mathcal{A}(\hat{x}^+_k)\|,$$
because $\|\mathcal{A}(\hat{x}^+_k)\| = \sqrt{a_{1k}^2 + a_{2k}^2+\dots+a_{mk}^2}$. This gives us the bound of $\|G_k\|^2$,
\[\|G_k\|^2 \le \delta \mathbf{v}^2\| \tilde{\mathcal{A}}(\hat{x}^+_k)^2 \| \le \delta \mathbf{v}^2\|\mathcal{A}(\hat{x}^+_k) \|. \]

Our next step is to find bound of $\|\mathcal{A}(\hat{x}^+_k)\|$, with given information of system. We implement norm inequality,
\begin{gather*}
	\|\mathcal{A}(\hat{x}^+_k) \| - \|\mathcal{A}(\hat{x}^+_k) - \mathcal{A}(x_k) \| \le \|\mathcal{A}(x_k)\| \\
	\implies \|\mathcal{A}(\hat{x}^+_k) \| \le \|\mathcal{A}(\hat{x}^+_k) - \mathcal{A}(x_k) \| + \|\mathcal{A}(x_k)\|. \hspace*{15pt}
\end{gather*}
The bound of propensity vector norm is, \[C_A \ge \sup\limits_{x \in \Omega} \|\mathcal{A}(x)\|, \tag{5} \label{true_prop_bound}\] where $\Omega$ is the set of all possible state trajectories of the system and considered to be a compact invariant set. Naturally, $\|\mathcal{A}(x_k)\| \le C_A$. The final expression is, \[ \|\mathcal{A}(\hat{x}^+_k) \| \le L_a\|e_k\| + C_A, \tag{6} \label{prop_bound}\]
where $e_k = x_k - \hat{x}^+_k$.
%We can also find the lower bound of $L_a$.Rewriting the Lipschitz continuity in form of CLE from~\eqref{cle-eqn}: 
%\begin{gather*}
%	 \|x_k + \delta V\mathcal{A}(x_k) - \hat{x}^+_k - \delta V\mathcal{A}(\hat{x}^+_k) \| \le L_f\|x_k - \hat{x}^+_k\| \\
%	 \implies \|e_k -\delta V(\mathcal{A}(\hat{x}^+_k) - \mathcal{A}(x_k)) \| \le L_f\|e_k\| \hspace*{1cm} \\
%	\implies \|e_k\| - \delta\mathbf{v}\|\mathcal{A}(\hat{x}^+_k) - \mathcal{A}(x_k) \| \le L_f\|e_k\| \hspace*{1cm} \\
%	\implies (1 - L_f)\|e_k\| \le \delta\mathbf{v}\|\mathcal{A}(\hat{x}^+_k) - \mathcal{A}(x_k) \| \hspace*{1.2cm} \\
%	\implies \frac{1-L_f}{\delta\mathbf{v}}\|e_k\| \le L_a\|e_k\| \implies L_a \ge \frac{1-L_f}{\delta \mathbf{v}} \tag{6} \label{la_lb}
%\end{gather*}
Using the final expression~\eqref{prop_bound}, bound of $\|G_k\|^2$:
\[ \|G_k\|^2 \le \delta \mathbf{v}^2C_A + \delta \mathbf{v}^2L_a\|e_k\|. \tag{7} \label{noise-bound-final} \]
\textbf{Step 2:} The filter error dynamics can be represented as,
\begin{gather*} 
	e_{k+1} = (I - K_{k+1}C)(f(x_k) - f(\hat{x}^+_k) + G_kw_k)  - K_{k+1}v_{k+1} \\= e_{kp} - e_{km}. \hspace*{6.5cm}
	\tag{8} \label{err-eqn} 
	\end{gather*}
$e_{kp}$ represents the error due to system dynamics and $e_{km}$ is error due to measurement noise. Since we need to find stability in mean square sense, we need to compute $E(\|e_{k+1}\|^2)$.

We initially find the bound of $\|P_{k+1}^+\|$, to eventually find $\|K_{k+1}\|$, $\|I - K_{k+1}C\|$ bound. It has been proved in Theorem 1.(i)~\cite{wei2025stochastic} that,
\[ \|P_{k+1}^+\| \le L_f^2\|P_k^+\| + \|G_k\|^2. \tag{9} \label{cov-bound} \]
Substituting value of $\|G_k\|^2$ in~\eqref{noise-bound-final}:
\[ \|P_{k+1}^+\| \le L_f^2\|P_k^+\| + \delta \mathbf{v}^2C_A + \delta \mathbf{v}^2L_a\|e_k\|. \]
Recursive substitution of $\|P_k^+\|$ leads to the expression,
\begin{align*}
	 \|P_{k+1}^+\| \le L^{2(k+1)}_f\|P_0^+\| + \delta\mathbf{v}^2C_A(1 + L_f^2 + L_f^4 \dots + L_f^{2k}) + \delta \mathbf{v}^2L_a(\|e_k\| + \\ L_f^2\|e_{k-1}\| + L_f^4\|e_{k-2}\| + \dots +L_f^{2k}\|e_0\|). \hspace*{5cm}
\end{align*}
Since $L_f < 1$, along with other known information $e_0, P_0^+$, for $k >> 1$, the final expression is obtained, 
\begin{gather*}
	\|P_{k+1}^+\| \le L^{2(k+1)}_f\|P_0^+\| + \delta \mathbf{v}^2L_a\|e_k\| + \frac{\delta\mathbf{v}^2C_A}{1 - L_f^2}
	\le \delta\beta_1 + \delta\beta_0\|e_k\|.  \tag{10} \label{err-final-exp}
\end{gather*}
Here, $\beta_0 = \mathbf{v}^2L_a$ and $\beta_1 = \frac{\mathbf{v}^2C_A}{1 - L_f^2}$.  Using expression of~\eqref{err-final-exp}, we can further obtain $\|K_{k+1}\|$ where $\|C\| = \mathbf{c}$ and $\rho = \frac{\mathbf{c}}{r}$,
\begin{gather*}
	\|K_{k+1}\| \le \|P_{k+1}^-C^T\|\|(CP_{k+1}^-C^T + R)^{-1}\|
\end{gather*}
As shown in proof of Theorem1.(ii)~\cite{wei2025stochastic} $\|(CP_{k+1}^-C^T + R)^{-1}\| < \|R^{-1}\|$, the inequality formed is,
\begin{gather*}
	\|K_{k+1}\| \le \|P_{k+1}^-\|\|C\|\|R^{-1}\| \le \left\{\frac{\mathbf{c}}{r} \|P_{k+1}^-\|\right\}
	\le \rho\delta\beta_1 \\+ \rho \delta\beta_0\|e_k\|. \hspace*{5.5cm} \tag{11} \label{k-norm-exp} 
\end{gather*}
We use $\|K_{k+1}\|$ to obtain bound of $\|I-K_{k+1}C\|$,
\begin{gather*}
	\|I-K_{k+1}C\| \le 1 + \|P_{k+1}^+C^TR^{-1}C\| \le 1 + \rho\delta\beta_1\mathbf{c} + \delta\beta_0\rho\mathbf{c}\|e_k\|
	 \\ \le \alpha + \delta\beta\|e_k\|. \hspace*{6.7cm} \tag{12} \label{err-gain-exp}
\end{gather*}
Here $\alpha = 1 + \rho\mathbf{c}\delta\beta_1$ and $\beta = \beta_0\rho\mathbf{c}$. The bound of $\|e_{k+1}\|^2$ is calculated  as,
\begin{gather*}
	\|e_{k+1}\|^2 \le \|(I-K_{k+1}C)(f(x_k) - f(\hat{x}^+_k) + G_kw_k)\|^2 + \|K_{k+1}v_{k+1}\|^2 \\ \le \|(I-K_{k+1}C)(\Delta f + G_kw_k)\|^2 + \|K_{k+1}v_{k+1}\|^2. \hspace*{3cm}
\end{gather*}
\textbf{Step 3:} The crucial point of this step is to find behavior of estimate error in next time instant given past estimate error values are known. We find $E(\|e_{k+1}\|^2 | e_k)$ expression as,
\begin{align*}
	E(\|e_{k+1}\|^2 | e_k) = E(\|e_{kp}\|^2 + \|e_{km}\|^2 - e_{kp}^Te_{km} - e_{km}^Te_{kp})
\end{align*}
Since $e_{km}$ contain $v_{k+1}$ term which is zero mean distribution, $E(v_{k+1}) = 0$. Hence $E(\|e_{k+1}\|^2 | e_k)$ is bounded by,
\begin{align*} 
	E(\|e_{k+1}\|^2 | e_k) \le E(\|e_{kp}\|^2|e_k) + E(\|e_{km}\|^2 | e_k). \hspace*{4mm}
\end{align*}
Using norm inequality properties with $E(w_k) = 0$, $E(\|w_k\|^2) = m$, $E(\|v_{k+1}\|^2) = pr$ and~\eqref{err-gain-exp},
\begin{gather*}
	E(\|e_{kp}\|^2|e_k) \le E((\alpha + \delta\beta\|e_k\|)^2(L_f^2\|e_k\|^2 + \Delta f^TG_kw_k +  w_k^TG_k^T\Delta f + \|G_k\|^2\|w_k\|^2)) \hspace*{3.5cm} \\
	 \le \alpha^2L_f^2\|e_k\|^2 + \delta^2\beta^2L_f^2\|e_k\|^4 + 2\delta\alpha\beta L_f^2\|e_k\|^3 + (\alpha^2 + \delta^2\beta^2\|e_k\|^2 + 2\delta\alpha \beta\|e_k\|)(\delta\mathbf{v}^2C_A  \\+ \delta\beta_0\|e_k\|)E(\|w_k\|^2)
	\le \alpha^2L_f^2\|e_k\|^2 + \delta^2\beta^2L_f^2\|e_k\|^4 + 2\delta\alpha\beta L_f^2\|e_k\|^3 + m\delta\alpha^2\mathbf{v}^2C_A + \\ m\delta^3\beta^2\mathbf{v}^2C_A\|e_k\|^2 +  m\delta^3\beta^2\beta_0\|e_k\|^3 +  2m\delta^2\mathbf{v}^2C_A\alpha\beta\|e_k\| + 2m \delta^2\beta_0\alpha\beta\|e_k\|^2 + m\delta\alpha^2\beta_0\|e_k\|,
\end{gather*}
and,
\begin{gather*}
	E(\|e_{km}\|^2 | e_k) \le (\rho\delta\beta_1 + \delta\beta_0\rho\|e_k\|)^2E(\|v_{k+1}\|^2)	\le (\rho\delta\beta_1)^2pr + (\delta\beta_0\rho)^2 pr\|e_k\|^2  \\+ 2\delta^2\rho^2\beta_1\beta_0pr\|e_k\|
	\equiv \delta^2\beta_1^2\rho\mathbf{c}p + \delta^2\beta_0\beta p\|e_k\|^2 + 2\delta^2\beta_1\beta p\|e_k\|. \hspace*{1cm}
\end{gather*}
The final expression is, 
\begin{gather*}
	E(\|e_{k+1}\|^2 | e_k) \le (\alpha^2L_f^2 + m\delta^3\mathbf{v}^2C_A\beta^2 + \delta^2\beta_0\beta(2m\alpha + p))\|e_k\|^2 + \delta^2\beta^2L_f^2\|e_k\|^4 + \\ \delta\beta(2\alpha L_f^2 + m\delta^2\beta_0\beta)\|e_k\|^3 + (m\alpha^2\delta\beta_0 + 2\delta^2\beta(m\alpha\mathbf{v}^2C_A 
	  + \beta_1p))\|e_k\| + \delta^2\beta_1^2p\rho\mathbf{c} +\\ m\alpha^2\delta\mathbf{v}^2C_A, \hspace*{9cm} \tag{13} \label{cond-exp}
\end{gather*}

Implementing $E(X) = E(E(X|Y))$
\begin{gather*}
	E(\|e_{k+1}\|^2) \le (\alpha^2L_f^2 + m\delta^3\beta^2\mathbf{v}^2C_A + (2m\alpha + p)\delta^2\beta_0\beta)E(\|e_k\|^2) + \delta^2\beta^2L_f^2E(\|e_k\|^4) \\+ (2\alpha L_f^2 + m\delta^2\beta_0\beta)\delta\beta E(\|e_k\|^3) + (m\alpha^2\delta\beta_0 + 2\delta^2\beta(m\alpha\mathbf{v}^2C_A 
	+ \beta_1p))E(\|e_k\|) + \\ m\alpha^2\delta\mathbf{v}^2C_A + \delta^2\beta_1^2p\rho\mathbf{c}. \hspace*{7.5cm} \tag{14} \label{mse_bound}
\end{gather*}

\textbf{Step 4:} The higher order moments of $\|e_k\|$ can be represented in terms of initial error covariance eigenvalues. Let $tr(P^+_k)$ denote trace of posterior error covariance matrix and we know that $tr(P^+_k) = \sum_{i=1}^{n}\lambda_{i,k}$ where $\lambda_{i,k}$ denotes $i^{th}$ eigenvalue of $k^{th}$ instant. It can also be related with $\|e_k\|^2$,
\[ tr(P^+_k) = E(\|e_k\|^2) \]
The variance formula if $\|e_k\|$ is a random variable can be used to get, 
\[ E(\|e_k\|^2) = var(\|e_k\|) + (E(\|e_k\|))^2 \implies E(\|e_k\|) < \sqrt{\sum_{i=1}^{n}\lambda_{i,k}}\]
We can use~\eqref{err-final-exp},
\begin{gather*}
	\sum_{i=1}^{n}\lambda_{i,k} \le n\lambda_{max}(P^+_k) \equiv n\|P^+_k\| \implies
	\sum_{i=1}^{n}\lambda_{i,k} \le c_{\beta_1} + c_{\beta_0}\|e_{k-1}\|.
\end{gather*}
Here, $c_{\beta_0} = \delta n \beta_0$ and $c_{\beta_1} = \delta n\beta_1$. Applying expectation on left hand side does not change the value as mentioned eigenvalues are not random variables. Hence,
\begin{gather*}	\sum_{i=1}^{n}\lambda_{i,k} \le c_{\beta_1} + c_{\beta_0}E(\|e_{k-1}\|) \implies
	\sum_{i=1}^{n}\lambda_{i,k} \le c_{\beta_1} + c_{\beta_0}\sqrt{\sum_{i=1}^{n}\lambda_{i,k-1}}. \tag{15} \label{eig_exp}
\end{gather*}
Recursively substituting the eigenvalue sum and applying the inequality $\sqrt{a+b} \le \sqrt{a} + \sqrt{b}$,
\[ \sum_{i=1}^{n}\lambda_{i,k} \le c_{\beta_0}^2\left(\frac{\sum_{i=1}^{n}\lambda_{i,0}}{c_{\beta_0}^2}\right)^{\frac{1}{2^k}} + \sum_{j=0}^{k}c_{\beta_0}^2\left(\frac{c_{\beta_1}}{c_{\beta_0}^2}\right)^{\frac{1}{2^j}}. \tag{16} \label{eig_exp_mod} \]
$\sum_{i=1}^{n}\lambda_{i,0}$ represents the trace of initial posterior covariance matrix. When $k >> 1$, the expression simplifies to,
\[ \sum_{i=1}^{n}\lambda_{i,k} \le c_{\beta_0}^2 + c_{\beta_0}^2\sum_{j=0}^{k}\left(\frac{c_{\beta_1}}{c_{\beta_0}^2}\right)^{\frac{1}{2^j}}. \tag{17} \label{eig_exp_smfd} \]

\textbf{Remark: }In~\eqref{eig_exp} $c_{\beta_0}$, $c_{\beta_1}$ are constants and $\sum_{i=1}^{n}\lambda_{i,0}$ obtained in~\eqref{eig_exp_mod} is finite. Since the analysis is performed over the finite simulation and filtering horizon $k=0,1\dots N$, there exists a finite constant $s_{\lambda_k}$, such that the summation term on right hand side in~\eqref{eig_exp_smfd} is bounded, 
\[ \sum_{j=0}^{k}\left(\frac{c_{\beta_1}}{c_{\beta_0}^2}\right)^{\frac{1}{2^j}} \le s_{\lambda_k}. \]
So,
\[ \sum_{i=1}^{n}\lambda_{i,k} \le c_{\lambda}, \hspace*{1cm} c_{\lambda} = c_{\beta_0}^2(1 + s_{\lambda_k}) \tag{18} \label{eig_exp_final}\] 
The error dynamics in~\eqref{err-eqn} consists of $w_k$, $v_{k+1}$ which are uncorrelated Gaussian noises, so $e_k$ can be considered to have Gaussian distribution, i.e $e_k \sim \mathcal{N}(0,P_k^+)$. We can find the distribution nature of $\|e_k\|^2 \equiv e_k^Te_k$. Considering $P_k^+ > 0$ and has spectral decomposition, we use Corollary 2.1 in~\cite{dik1985distribution}, 
\[\|e_k\|^2 \sim \sum_{i=1}^{n}\lambda_{i,k}\mathcal{X}^2_{1i}\]
Using moment generating function of the above distribution, we find second and third order moment as,
\begin{gather*}
	E(\|e_k\|^4) = 2\sum_{i=1}^{n}\lambda_{i,k}^2 + \left(\sum_{i=1}^{n}\lambda_{i,k} \right)^2, \hspace*{7cm}\\
	E(\|e_k\|^6) = 8\sum_{i=1}^{n}\lambda_{i,k}^3 + 6\sum_{i=1}^{n}\lambda_{i,k}\sum_{i=1}^{n}\lambda_{i,k}^2 + \left(\sum_{i=1}^{n}\lambda_{i,k} \right)^3 \hspace*{4cm}.
\end{gather*}
To find $E(\|e_k\|^3)$, we use $E(\|e_k\|^6) = var(\|e_k\|^3) + (E(\|e_k\|^3))^2$ which gives $E(\|e_k\|^3) <  \sqrt{E(\|e_k\|^6)}$,
\[ \implies E(\|e_k\|^3) < \sqrt{8\sum_{i=1}^{n}\lambda_{i,k}^3} + \sqrt{6\sum_{i=1}^{n}\lambda_{i,k}\sum_{i=1}^{n}\lambda_{i,k}^2} + \sqrt{\left(\sum_{i=1}^{n}\lambda_{i,k} \right)^3} \]
We can substitute $\sum_{i=1}^{n}\lambda_{i,k}^3 < (\sum_{i=1}^{n}\lambda_{i,k})^3$ and $\sum_{i=1}^{n}\lambda_{i,k}^2 < (\sum_{i=1}^{n}\lambda_{i,k})^2$ getting all moments in terms of $tr(P^+_k)$,

\begin{gather*}
	E(\|e_k\|^3) < 3.83\left(\sum_{i=1}^{n}\lambda_{i,k}\right)^3 + 2.45\sqrt{\sum_{i=1}^{n}\lambda_{i,k}}\sum_{i=1}^{n}\lambda_{i,k} \equiv 3.83c_{\lambda}^3 + 2.45c_{\lambda}^{\frac{3}{2}} \\
	E(\|e_k\|^4) < 3\left(\sum_{i=1}^{n}\lambda_{i,k}\right)^2 \equiv 3c_{\lambda}^2 \hspace*{6cm}
\end{gather*}
Substituting all moments except $E(\|e_k\|^2)$, 
\begin{gather*}
	E(\|e_{k+1}\|^2) \le (\alpha^2L_f^2 + m\delta^3\beta^2\mathbf{v}^2C_A + (2m\alpha + p)\delta^2\beta_0\beta)E(\|e_k\|^2) + 3\delta^2\beta^2L_f^2c_{\lambda}^2 \\+ \delta\beta(2\alpha L_f^2 + m\delta^2\beta_0\beta)(3.83c_{\lambda}^3 + 2.45c_{\lambda}^{\frac{3}{2}}) + (m\alpha^2\delta\beta_0 + 2\delta^2\beta(m\alpha\mathbf{v}^2C_A 
	+ \beta_1p))c_{\lambda}^{\frac{1}{2}} \\+ m\alpha^2\delta\mathbf{v}^2C_A + \delta^2\beta_1^2p\rho\mathbf{c}. \hspace*{7.3cm} \tag{19} \label{err-bound-exp}
\end{gather*}

As per Definition 1, the error can be exponentially bounded if $\gamma < 1$. In~\eqref{err-bound-exp}, coefficient of $E(\|e_k\|^2)$ the parameters are predetermined based on system/ process except $\delta$. Tuning $\delta$ to satisfy the conditions ensure boundedness of the process is maintained. Also, $\gamma$ forms a polynomial function of $\delta$. Replacing $\alpha$ with its original term and substituting $c_2 = \beta_1\rho\mathbf{c}$, we get the polynomial inequality,
 \begin{gather*}
 	\delta^3(m\mathbf{v}^2C_A\beta^2 + 2mc_2\beta_0\beta + \delta^2(c_2^2L_f^2 + (2m +p)\beta_0\beta) + \delta 2c_2L_f^2 + L_f^2 - 1 < 0. \tag{20} \label{third_order_poly}
 \end{gather*} 
Based on Descartes rule of signs~\cite{wang2004simple}, the zero power term $L_f^2 -1$ has negative sign since $L_f <1$ based on Assumption 2, whereas higher order coefficient have positive signs as system defined parameters and roots of posterior error covariance are positive. Hence, only one positive root exists based on sign change. The threshold value of $\delta$ is positive root value, which ensures the exponential boundedness condition is satisfied. Using $\gamma < 1$ for appropriate $\delta$,
%Hence we make use of~\eqref{la_lb}, where we implement the minimum value of $L_a$ to reduce the above expression into a quadratic inequality with $\beta_0' = (1-L_f)\mathbf{v}$ and $\beta' = (1-L_f)\mathbf{v}\rho\mathbf{c}$:
%\begin{gather*}
%	\delta^2c_2^2L_f^2 + \delta(c_2((2 + m_1\beta')L_f^2 + 2m\beta_0'\beta') + m\mathbf{v}^2C_A\beta'^2) \\
%	+ L_f^2 + \beta_0'\beta'(2m + p) + \beta'^2L_f^2(0.25m_1^2 + 1.5m_2) - 1 +\\ \beta'L_f^2m_1+ 0.5mm_1\beta_0'\beta'^2  < 0 \hspace*{3cm} \tag{17} \label{del-quad-eqn}
%\end{gather*}
%We solve the quadratic equation to get roots of $\delta$ where,
%$$ \delta < \frac{-b \pm \sqrt{b^2 - 4ac}}{2a}, $$
%which ensures the exponential boundedness condition is satisfied. Using $\gamma < 1$ for appropriate $\delta$,
\begin{gather*}
	E(\|e_{k+1}\|^2) \le \gamma E(\|e_k\|^2) + C_0' 
	\le \gamma^{k+1}E(\|e_0\|^2) + \frac{C_0'}{1 - \gamma} \le \gamma^{k+1}E(\|e_0\|^2) + C_0.
\end{gather*}
The positive root of the polynomial in~\eqref{third_order_poly} can be found numerically.
\section{EXAMPLE AND SIMULATION}
We consider the basic gene expression~\cite{dash2025extended} to demonstrate the filter error performance with respect to time. The discrete system dynamics are given as,
\begin{gather*}
	P_{o_{k+1}} = P_{o_k} + \delta((k_{up} + k_{tx})(P_{tot} - P_{o_k}) - k_{bp}GP_{o_k}) + \sqrt{\delta} \left(\sqrt{k_{up}(P_{tot} - P_{o_k})}w_2 \right.\\ \left. +\sqrt{k_{tx}(P_{tot} - P_{o_k})}w_3  -\sqrt{k_{bp}GP_{o_k}}w_1\right) \hspace*{5.5cm}\\
	T_{k+1} = T_k + \delta(k_{tx}(P_{tot} - P_{o_k}) + (k_{tl} + k_{ur})(R_{tot} - R_{i_k})
	- k_{br}T_kR_{i_k} - d_{T}T_k)
	+\\ \sqrt{\delta}\left(\sqrt{k_{tx}(P_{tot} - P_{o_k})}w_3 -   \sqrt{k_{br}T_kR_{i_k}}w_4 + \sqrt{ k_{ur}(R_{tot} - R_{i_k})}w_5 - \sqrt{d_{T}T_k}w_7 \right. \\ \left. + \sqrt{k_{tl}(R_{tot} - R_{i_k})}w_6 \right) \hspace*{8cm}\\
	R_{i_{k+1}} = R_{i_k} + \delta((k_{ur} + k_{tl})(R_{tot} - R_{i_k}) - k_{br}T_kR_{i_k}) + \sqrt{\delta}\left( \sqrt{k_{tl}(R_{tot} - R_{i_k})}w_6  
	\right. \\ \left. + \sqrt{k_{ur}(R_{tot} - R_{i_k})}w_5 -  \sqrt{k_{br}T_kR_{i_k}}w_4\right) \hspace*{5.8cm}  \\
	X_{k+1} = X_k + \delta(k_{tl}(R_{tot} - R_{i_k}) - d_{X}X_k) + \sqrt{\delta}\left(\sqrt{k_{tl}(R_{tot} - R_{i_k})}w_6 - \sqrt{d_{X}X_k}w_8 \right). \tag{21} \label{cle-ge} 
\end{gather*}
\begin{figure}[thpb]
	\centering
	\begin{subfigure}{42mm}
		\includegraphics[width=45mm]{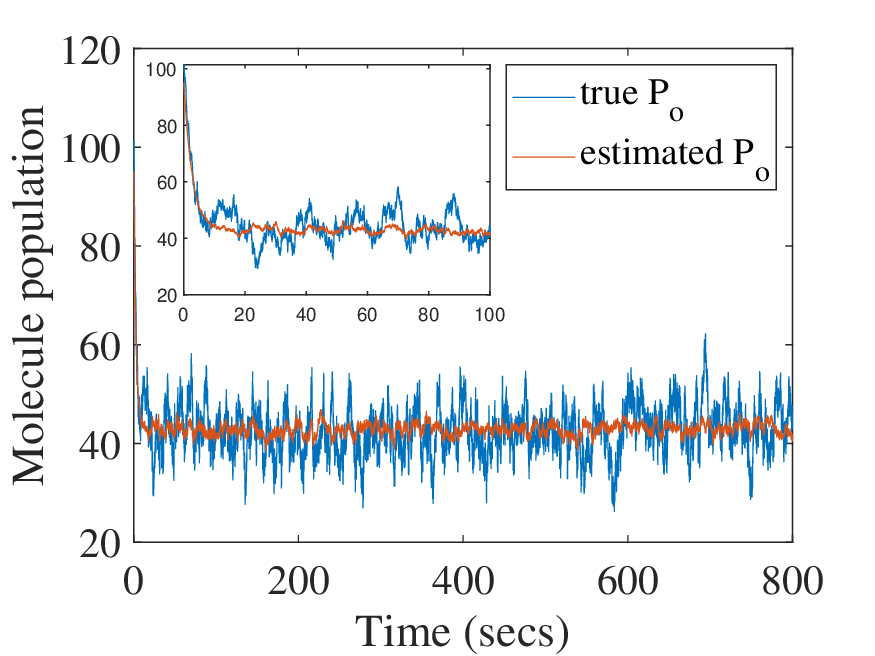}
		\subcaption{}
	\end{subfigure}
	\begin{subfigure}{42mm}
		\includegraphics[width=45mm]{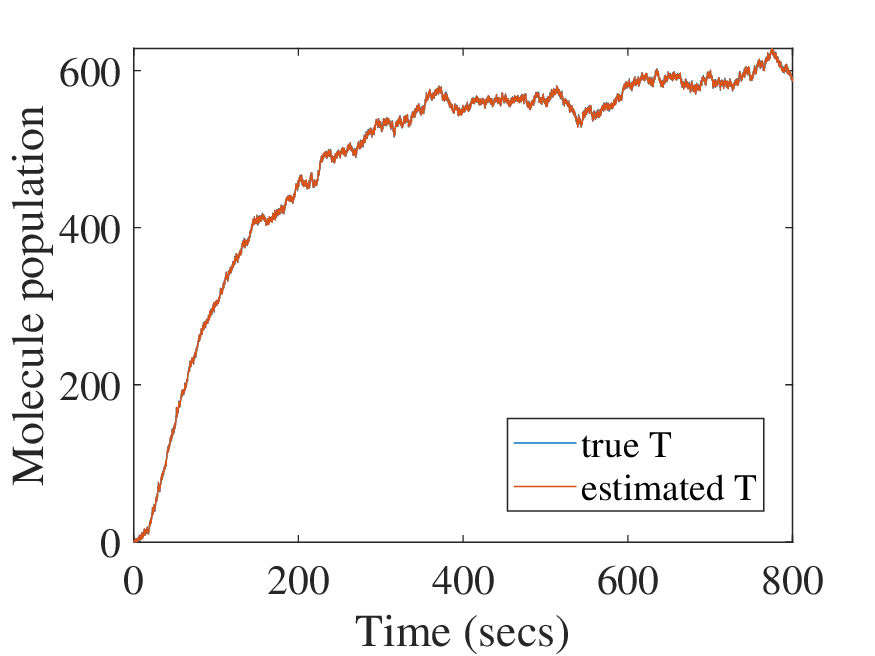}
		\subcaption{}
	\end{subfigure}
	\begin{subfigure}{42mm}
		\includegraphics[width=45mm]{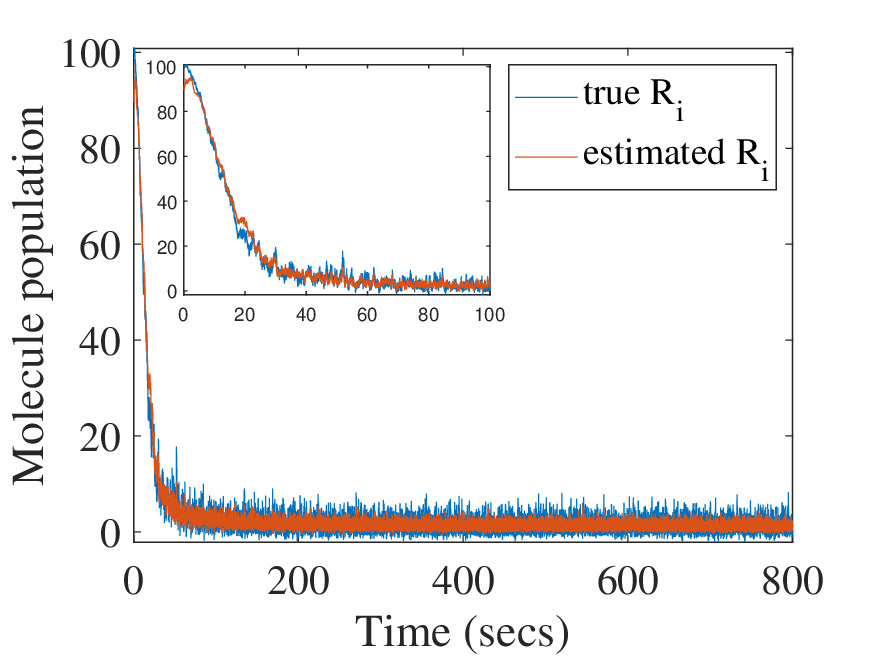}
		\subcaption{}
	\end{subfigure}
	\begin{subfigure}{42mm}
		\includegraphics[width=45mm]{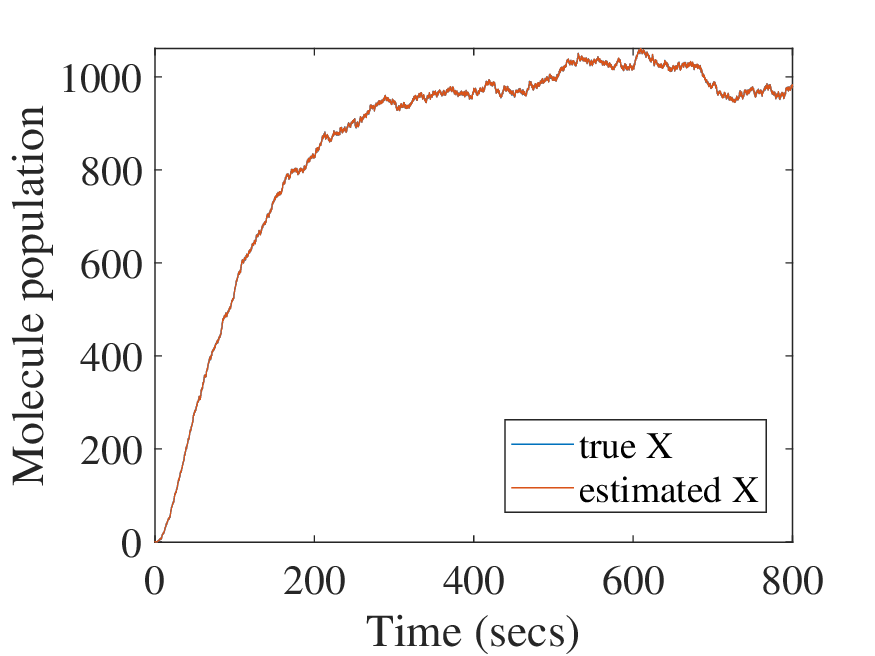}
		\subcaption{}
	\end{subfigure}
	\caption{Filter error obtained for $(a)$ RNA polymerase $(b)$ mRNA transcript $(c)$ ribosome and $(d)$ protein states}
	\label{est_error_states}
\end{figure}
Here, $P_o$ indicates the RNA polymerase, $T$ denotes the mRNA transcript, $R_i$ denotes ribosome, $X$ indicates protein biomolecules and $w_i$ represents the process noise of $i^{th}$ reaction of normal distribution $\mathcal{N}(0,1)$. Propensity vector and stochiometric matrix for the specific system is represented as, 
\begin{gather*}
	\mathcal{A}(x_k)=[
	k_{bp}GP_{o_k}, k_{up}(P_{tot}- P_{o_k}), k_{tx}(P_{tot} - P_{o_k}), \\ k_{br}T_kR_{i_k}, k_{ur}(R_{tot} - R_{i_k}),  k_{tl}(R_{tot} - R_{i_k}), d_TT_k, d_XX_k
	]^T, 
\end{gather*}
and,
\begin{gather*}
	V = \begin{bmatrix}
		-1 & 1 & 1 & 0 & 0 & 0 & 0 & 0 \\
		0 & 0 & 1 & -1 & 1 & 1 & -1 & 0\\
		0 & 0 & 0 & -1 & 1 & 1 & 0 & 0 \\
		0 & 0 & 0 & 0 & 0 & 1 & 0 & -1
	\end{bmatrix}.
\end{gather*}
Both $T$ and $X$ are measurable with same noise distribution meaning,  $$C = 
\begin{bmatrix}
	0 & 1 & 0 & 0 \\
	0 & 0 & 0 & 1
\end{bmatrix} \implies \mathbf{c} = 1$$.\newline
With this example, we try to find a suitable $\delta$ for which EKF with updated process noise covariance will be stable in a stochastic sense. Based on $V$ for the system, $\mathbf{v} = 2.7657$ with initial biomolecule population $P_{o_0} = 100$, $T_0 = 0$, $R_{i_0} = 100$, $X_0 = 0$, kinetic rates $k_{tx} = 0.1$, $k_{up} = 0.05$, $k_{bp} = 0.02$, $k_{tl} = 0.1$, $k_{ur} = 0.05$, $k_{br} = 0.02$, $d_T = 0.01$, $d_X = 0.01$ were considered for data generation. Measurement noise covariance is $rI = 10I_{2 \times 2}$ and measurements were generated by adding simulated data with multivariate normal distribution generator \textit{mvrnd}. Initial estimates are taken as $\hat{P_{o_0}} = 95$. $\hat{T}_0 = 0$, $\hat{R_{i_0}} = 90$ and $\hat{X}_0 = 0$ with initial error covariance matrix, 
\[ P^+_0 = \begin{bmatrix}
	23 & 0 & 0 & 0 \\
	0 & 1 & 0 & 0 \\
	0 & 0 & 45 & 0 \\
	0 & 0 & 0 & 1
\end{bmatrix}.\] 
We compute $C_A = 55$ empirically from Monte Carlo stochastic simulation of all the datasets considered using~\eqref{true_prop_bound}. Based on Assumptions 1 and 4, $L_f = 0.85$, $L_a = 0.8$. 
\begin{figure}[thpb]
	\centering
	\begin{subfigure}{42mm}
		\includegraphics[width=45mm]{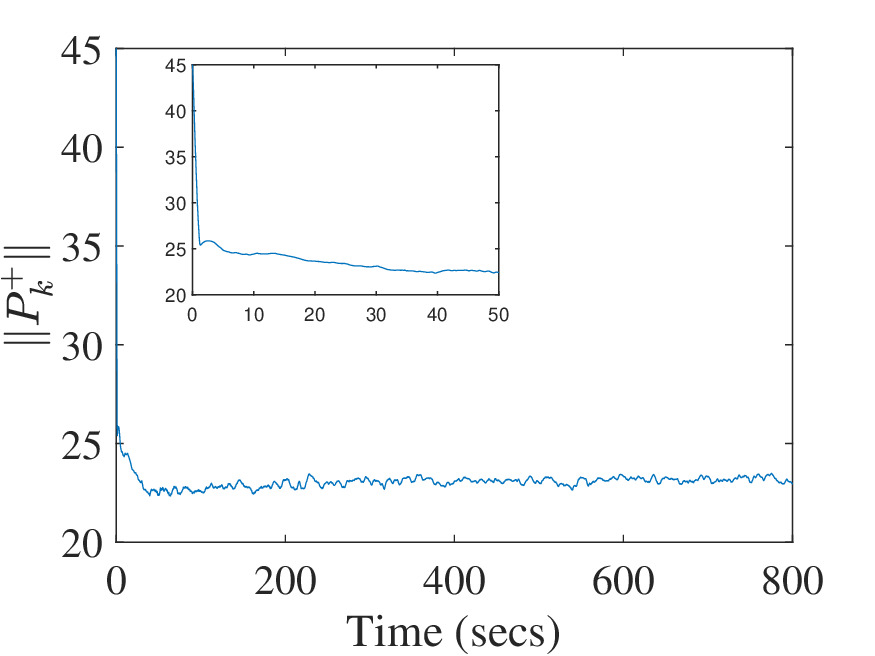}
		\subcaption{}
		\label{post_err_res}
	\end{subfigure}
	\begin{subfigure}{42mm}
		\includegraphics[width=45mm]{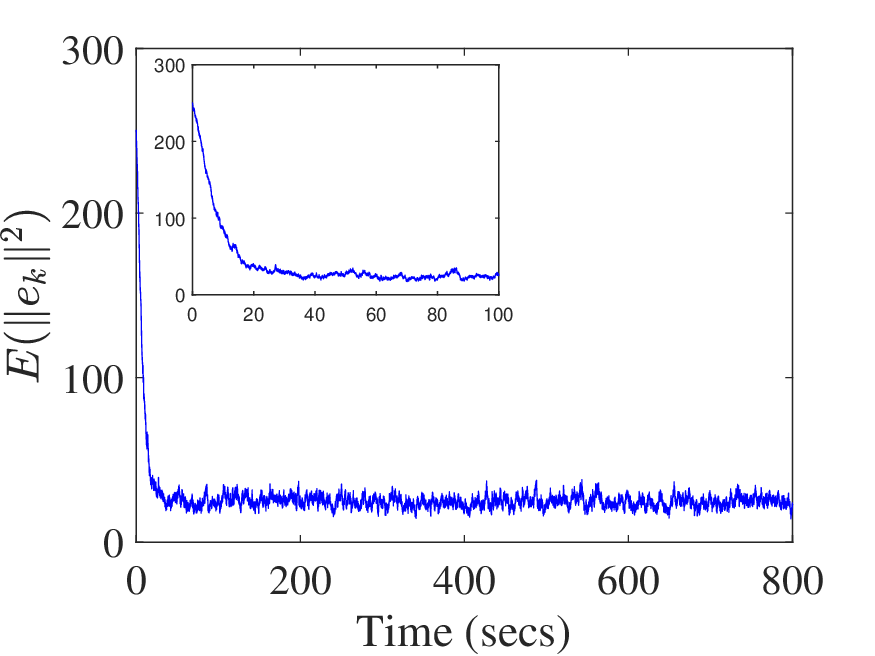}
		\subcaption{}
		\label{main_res}
	\end{subfigure}
	\caption{Temporal plot of $(a)$ posterior error covariance matrix norm $(b)$ mean square filter error norm}
\end{figure}
Using the above system based and assumed values, we calculate coefficients of inequality in~\eqref{third_order_poly} to be $10345.68$, $16672.97$, $219.07$ and $-0.28$ corresponding to descending order of power terms using the \textit{roots} function in MATLAB. The positive root obtained is $\delta_{max} = 0.0011$. As mentioned in Theorem 1, $\delta_{max}$ is a conservative bound as it is hard to exactly define $L_f$ and $L_a$ values. Based on aforementioned values, $100$ datasets were generated using Euler-Maruyama simulation for $800$ seconds with $\delta = 0.001$ as sampling interval. EKF with estimate dependent process noise covariance is implemented to obtain estimated states. As observed in~\cite{dash2025extended}, the temporal norm of process noise covariance matrix reaches a steady state in mean, confirming boundedness. We find the estimation error for all system states as shown in Fig.~\ref{est_error_states}.
Additionally, we check the characteristic of posterior error covariance matrix through its spectral norm shown in Fig.~\ref{post_err_res}. We check the filter error performance by comparing the true state generated and estimated state by the proposed filter. We observe that the estimate error covariance matrix represents an oscillatory bounded behavior in Fig.~\ref{main_res}. The metric $E(\|e_k\|^2)$ shows an exponentially decreasing behavior with noisy property, bounded by an upper limit. 

%We also compare behavior of $E(\|e_k\|^2)$ of updated process noise covariance matrix ($\hat{Q_k}$) with fixed Q for $150$ seconds. The results are shown in Fig.~\ref{q_comp_fig} where we observe in case of $\hat{Q_k}$, it shows a better behavior of exponential bound compared other fixed Q.
%\begin{figure}
%	\includegraphics[scale=0.6]{q_comp_err}
%	\caption{Comparison of $E(\|e_k\|^2)$ for different Q till 150 seconds }
%	\label{q_comp_fig}
%\end{figure}
\section{CONCLUSION}
We validated the stochastic stability in exponential sense of an EKF with estimate dependent process noise covariance where exponentially bounded nature of filter error is observed for a certain range of sampling interval, dependent on system based parameters. We also verified our theory with discrete time gene expression system. This gives significant information on this particular adaptive filter stability. The proposed filter offers useful design practices for state and parameter estimation in chemical reaction networks, and can be further extended to other estimation techniques like Rao-Blackwellized particle filters.


%%%%%%%%%%%%%%%%%%%%%%%%%%%%%%%%%%%%%%%%%%%%%%%%%%%%%%%%%%%%%%%%%%%%%%%%%%%%%%%%

%%%%%%%%%%%%%%%%%%%%%%%%%%%%%%%%%%%%%%%%%%%%%%%%%%%%%%%%%%%%%%%%%%%%%%%%%%%%%%%%

%%%%%%%%%%%%%%%%%%%%%%%%%%%%%%%%%%%%%%%%%%%%%%%%%%%%%%%%%%%%%%%%%%%%%%%%%%%%%%%%
%\section*{APPENDIX}


%\section*{ACKNOWLEDGMENT}

%%%%%%%%%%%%%%%%%%%%%%%%%%%%%%%%%%%%%%%%%%%%%%%%%%%%%%%%%%%%%%%%%%%%%%%%%%%%%%%%

\bibliographystyle{ieeetr}
\bibliography{refs.bib}

\end{document}